\documentclass[twocolumn,prl,superscriptaddress,showpacs,byrevtex]{revtex4}
\usepackage{mathptmx,amssymb}
\usepackage{amsmath,ulem}

\usepackage{graphicx}
\usepackage{color}
\usepackage{epstopdf}

\begin{document}

\title{Fluctuations of electrical conductivity: a new source for astrophysical magnetic fields}

\author{F. P\'etr\'elis, A. Alexakis, C. Gissinger}
\affiliation{Laboratoire de Physique Statistique, Ecole Normale Sup\'erieure, CNRS, Universit\'e P. et M. Curie, Universit\'e Paris Diderot, Paris, France}

\begin{abstract} 

We consider the generation of magnetic field by the flow of a fluid for which the electrical conductivity  is nonuniform.  A new amplification mechanism is found which leads to dynamo action for flows much simpler than those considered so far. In particular, the fluctuations of the electrical conductivity provide a way to bypass anti-dynamo theorems. For astrophysical objects, we show through three-dimensional global numerical simulations that the temperature-driven fluctuations of the electrical conductivity  can amplify an otherwise decaying large scale equatorial dipolar field. This effect could play a role for the generation of  the unusually tilted magnetic field of the iced giants Neptune and Uranus.
\end{abstract}
\maketitle

The current explanation for the existence of magnetic field in astrophysical objects was given in 1919 by Larmor \cite{larmor}. The motion of an electrically conducting fluid amplifies a seed of magnetic field by induction: this is the dynamo instability.
Despite nearly hundred years of research, several questions remain open. 
One of the reasons is that for a flow to be dynamo active, it has to be complex enough.  

For instance, for a fluid with uniform physical properties, planar flows  cannot create magnetic fields \cite{2D}. This result together with other similar anti-dynamo theorems  \cite{moff}, severely constrain the structure of the flows that  can act as dynamos. Broadly speaking, both the flow and the resulting magnetic field must be complex enough. 

In an astrophysical object, considering the electrical conductivity $\sigma$ as a constant is  a very crude simplification. 
In most natural situations (liquid core of planetary dynamos, plasmas of stellar convection zones, galaxies),  the temperature $T$, the chemical compositions $C_i$ and the density of the fluid  $\rho$ are expected to display  large variations. As a result the electrical conductivity of the fluid is unlikely to remain uniform in the bulk of the flow. 
In other words, $\sigma$, that is determined by $\rho$, $T$ and $C_i$ can be written as a function of space and time $\sigma(r,t)$ because $\rho$, $T$ and $C_i$ are functions of space and time.  
The effect of a boundary of varying conductivity close to a  flow tangent to the boundary had been considered to model  inhomogeneities of the Earth mantle \cite{busseandwicht}. A dynamo instability has been predicted but requires a flow with a huge velocity \cite{gallet}.  
 In this article we describe how magnetic field generation is affected by  conductivity variations in the bulk of the fluid.

To calculate this effect, we have to take into account that $\sigma$ depends on  position in the equation for the magnetic field that reads 
\begin{equation}
\frac{\partial {\bf B}}{\partial t}=\nabla \times ({\bf v} \times {\bf B}) - \nabla \times \left(\frac{1}{\sigma}\nabla \times ( \frac{{\bf B}}{\mu_0}) \right) 
\label{Eqbase}
\end{equation}
Insight can be  obtained  using the approximation of scale separation. We assume that the velocity and conductivity fields are periodic of period $l$. We note $\langle \cdot \rangle$ the spatial average over $l$. Let the magnetic diffusivity be $\eta=(\mu_0 \sigma)^{-1}=\eta_0+\delta \eta$, where $\eta_0$ is the mean of $\eta$ and  $\delta \eta$ its variations.  We write ${\bf B}=\langle {\bf B}  \rangle+ {\bf b}$ and consider that $\langle {\bf B}  \rangle$ varies on a very large scale  compared to $l$. In this limit, $\langle {\bf B}  \rangle$ satisfies a mean-field (closed) equation that reads
\begin{equation}
\frac{\partial \langle {\bf B} \rangle }{\partial t}=\nabla \times \left( \alpha \langle {\bf B}  \rangle \right) +\eta_0 \nabla^2 \langle {\bf B} \rangle\,.
\label{EqbaseB}
\end{equation}
where $\alpha \langle {\bf B}  \rangle$ is the sum of two terms, 
\begin{equation}
\alpha \langle {\bf B}  \rangle=\langle {\bf v} \times {\bf b}\rangle -  \langle \delta \eta \nabla \times {\bf b} \rangle.
\label{Eqdefal}
\end{equation}
Provided that $\delta \eta$  and the small scale field are small compared to respectively $\eta_0$ and the large scale field,  ${\bf b}$ is solution of
 \begin{equation}
\frac{\partial {\bf b}  }{\partial t}-\eta_0 \nabla^2 {\bf b}=\langle {\bf B}  \rangle \cdot \nabla {\bf v}\,,
\label{Eqbaseb}
 \end{equation}
such that by virtue of scale separation ${\bf b}$  can be calculated as a function of the large scale field $\langle {\bf B} \rangle$. Then  $\alpha$ is obtained  which closes  equation (\ref{EqbaseB}).  
The  term $\langle {\bf v} \times {\bf b}\rangle $ is the usual alpha-effect  \cite{moff} and writes $\langle {\bf v} \times {\bf b}\rangle=\alpha^{h} \langle {\bf B}\rangle  $. The tensor $\alpha^h$ can be expressed using the Fourier transform of the velocity field $\hat{\bf v}=(2\pi)^{-3/2}\int {\bf v} \exp (i {\bf k} r) d^3r$ where for simplicity we have set $l=2 \pi$ in all directions. We obtain
\begin{equation}
\alpha^h_{u,j}= (2 \pi)^{-3} i \Sigma_k \frac{{\bf k}_j}{\eta_0 {\bf k}^2} \left(\hat{{\bf v}}(-{\bf k})\times \hat{{\bf v}}({\bf k})\right)_u\,.\end{equation}
This is the usual result for the $\alpha$-tensor in an homogeneous fluid. 
The second term in Eq. \ref{Eqdefal} is new and reads
\begin{equation}
\alpha^{\sigma}_{u,j}  \langle {\bf B}_j  \rangle=-\langle \delta \eta \nabla \times {\bf b} \rangle  =(2 \pi)^{-3}  \Sigma_k \frac{{\bf k}. \langle {\bf B} \rangle}{\eta_0 {\bf k}^2} \widehat{\delta \eta}(-{\bf k}) \left({\bf k}\times \hat{{\bf v}}({\bf k})\right)_u\,.
\end{equation}
Introducing the vorticity ${\Omega}=\nabla \times {\bf v}$, the new part  of the $\alpha$-tensor can be written 
\begin{eqnarray}
\alpha^{\sigma}_{u,j}&=& -(2 \pi)^{-3} i \Sigma_k \frac{{\bf k}_j}{\eta_0 {\bf k}^2} \left(\widehat{\delta \eta}(-{\bf k}) \widehat{{ \Omega}}_u({\bf k})\right) \nonumber \\ 
&=&(2 \pi)^{-3} \Sigma_k \frac{\widehat{\partial_j \delta \eta}(-{\bf k}) \widehat{{ \Omega}}_u({\bf k})}{\eta_0 {\bf k}^2}\nonumber \\ &=&-(2 \pi)^{-3} \Sigma_k \frac{\widehat{\delta \eta}(-{\bf k}) \widehat{\partial_j{\Omega}}_u({\bf k})}{\eta_0 {\bf k}^2} 
\label{eqalsigma}
\end{eqnarray}

Large values of  $\alpha^{\sigma}$ thus require strong correlations between diffusivity variations and gradients of the vorticity or, equivalently, between gradients of diffusivity and vorticity. 
This can be understood by considering  
a vortical flow in which the vorticity is modulated in the $\phi$-direction, a classical picture of convective flows in planetary cores, as sketched in fig. (1). Assume that a large scale magnetic field is applied in the $\phi$-direction. Calculating ${\bf v}\times {\bf B}$, we observe that currents of opposite signs are induced in the vertical $z$-direction. Then, the azimuthal variation of electrical conductivity strengthens the current in one direction and reduces it in the opposite one. This results in a total electric current flowing in the $z$-direction as predicted by our calculation. This current can in turn amplify the magnetic field.

\begin{figure}[htb!]
\begin{center}
\includegraphics[width=80mm]{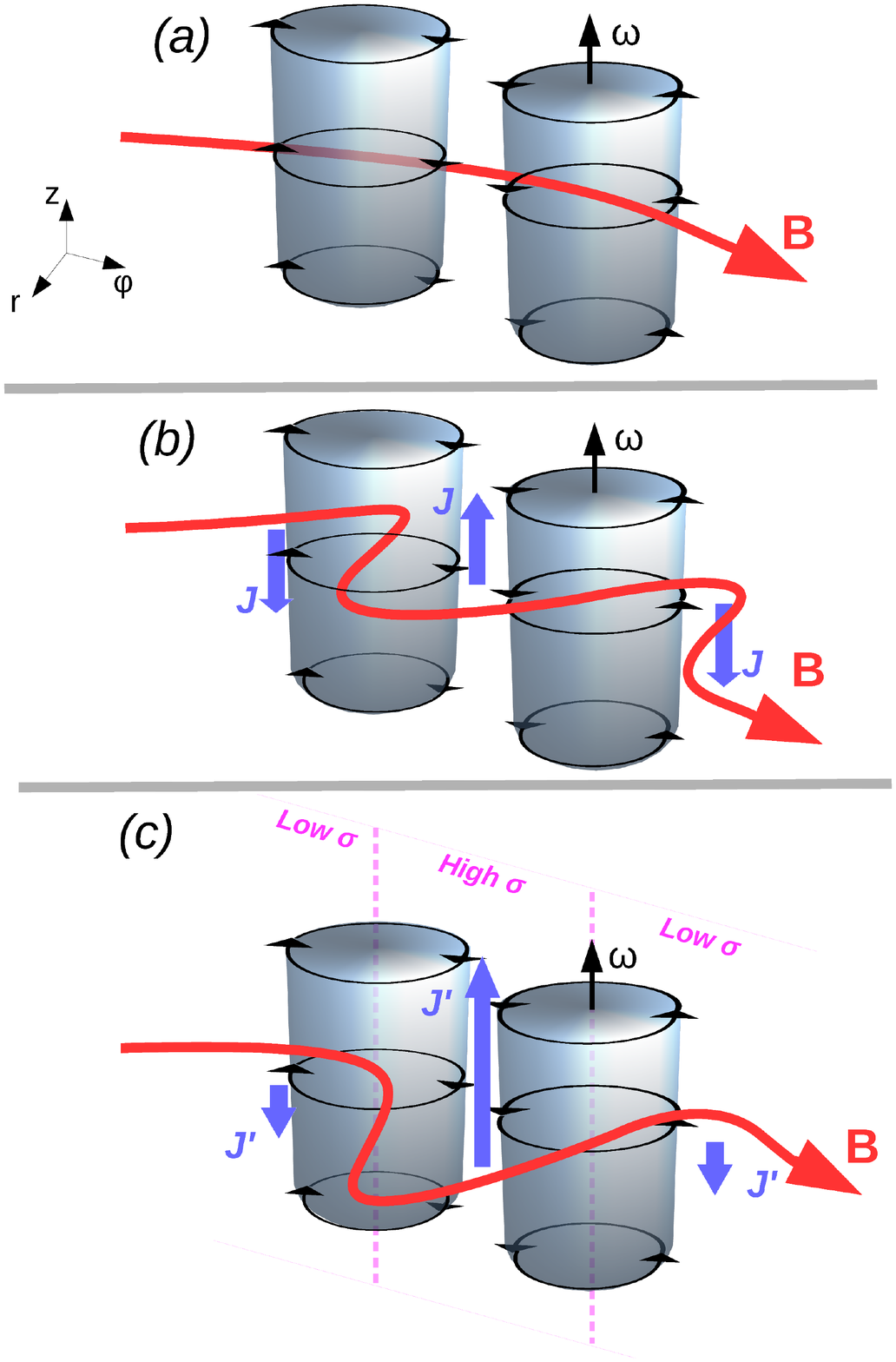}
\caption{Sketch of the different steps involved in the amplification mechanism $\alpha^{\sigma}$ for a typical geophysical flow. Top: Two adjacent convective cells (grey cylinders) with axial vorticity $\omega$ are subject to a transverse azimuthal magnetic field  $B$ (red). Middle: Both upward and downward axial currents $J\propto ({\bf v \times  B})$ (blue) are induced between the convective cells. Bottom: In presence of conductivity gradients correlated to the vorticity (maximum gradient  represented by pink dashed lines), large (resp. low) conductivity increases (resp. decreases) the induced current: the resulting net upward current $J^{'}$ is parallel to the vorticity. }
\label{fig2}
\end{center}
\end{figure}

Having identified the pertinent properties of the velocity and conductivity fields, we now discuss one example. 
Let the velocity be  
${\bf v}=(A \cos(k y) \sin(k z), B \cos(k x) \sin(k z), 0)$
 and  the diffusivity variation be 
$\delta \eta/\eta_0= \delta (\cos(k z) (\sin(k y)-\sin(k x)))$. The velocity field is a periodic array of counter-rotating vortices located in the x-y planes. The amplitude of the velocity field is simply modulated in the $z$-direction.  The $\alpha^{\sigma}$ tensor reads $\langle {\bf v} \times {\bf b} \rangle=0$ and $\langle-\delta \eta \nabla \times  {\bf b}\rangle=\delta/8\, (B Bx,\, A By,\, -(A+B) Bz)$.
We then calculate the growth rate $p$ for a large scale mode proportional to $\exp{(p t+i K z)}$ and obtain    $p=\frac{|\delta K| \sqrt{A B }}{8} - \eta_0 K^2$. 
Dynamo instability is possible provided $Rm =|\delta| \sqrt{A B }/(\eta_0 |K|)>8$. We point out that for this flow, in the absence of conductivity variation, no dynamo would be possible. 

The asymptotic results derived here were confirmed
using numerical simulations.  To achieve large scale separation,  we used a code based on Floquet theory, allowing us to write the solutions of Eq. (1) as
${\bf B}({\bf x},t)=e^{i \bf K\cdot x} {\bf b}({\bf x},t)$,
where ${\bf K}$ is an arbitrary  wavenumber
and ${\bf b}({\bf x},t)$ is a space-periodic vector field
with the same period as ${\bf v}$ and $\eta$.
The numerically calculated growth rates for the flow 
are shown in figure 2  for $Rm=1$ and different values of ${\bf K}$ and $\delta\eta$, and show an excellent agreement with the asymptotic results. Note that, because of scale separation, even small values of the diffusivity variation $\delta\eta$  lead to a dynamo. 

\begin{figure}[htb!]
\begin{center}
\includegraphics[width=90mm]{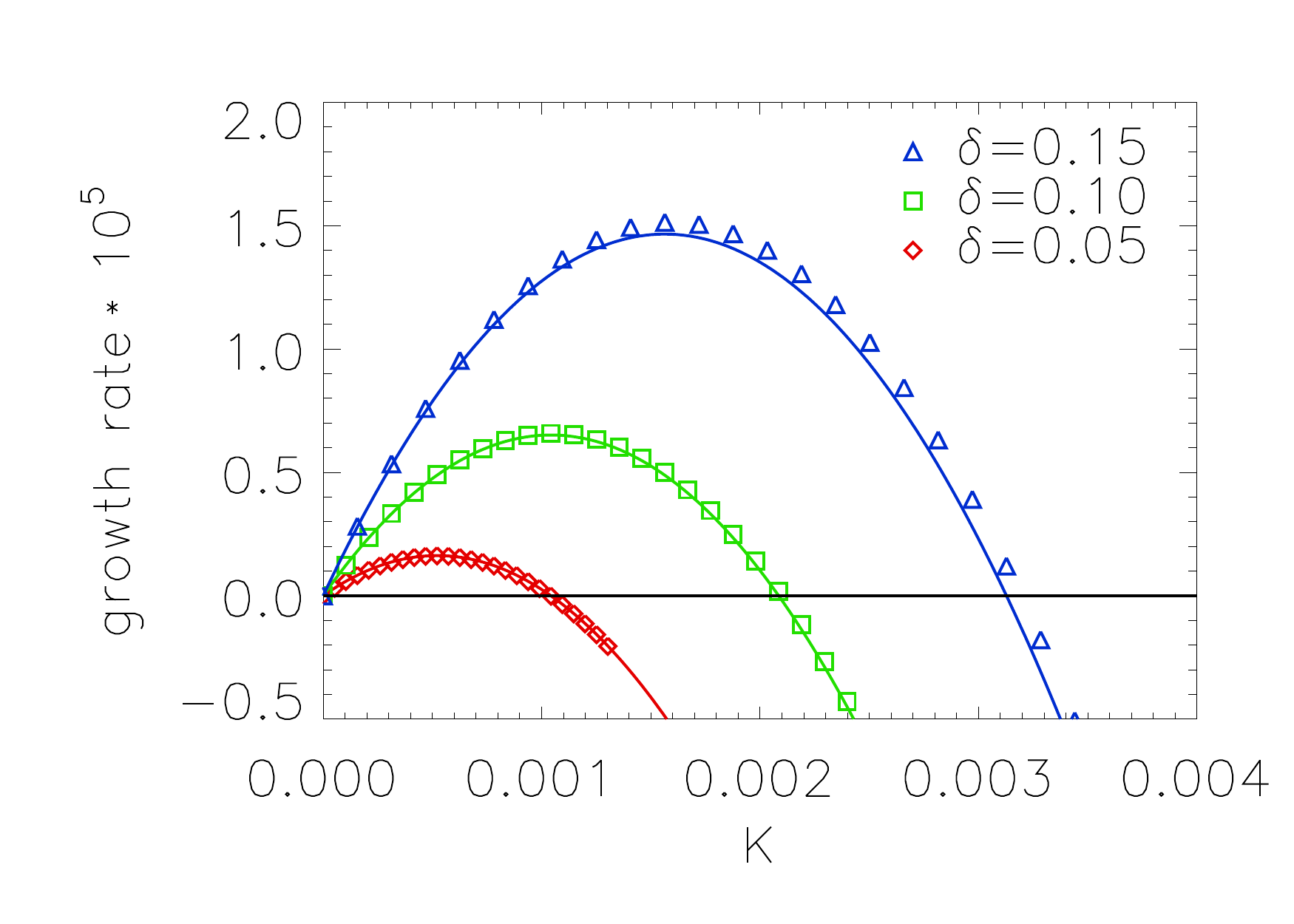}
\caption{
The growth rates for the 2D flow considered in the text as a function of   ${\bf K}$, for 
$Rm=1/6$  and three different $\delta \eta$.
Symbols correspond to numerically evaluated growth rates and straight 
lines to the analytical prediction.
}
\label{figalex}
\end{center}
\end{figure}

This mechanism provides a simple way to bypass anti-dynamo theorems and may thus play a  role in the creation of magnetic fields of astrophysical objects. 
As a first step towards answering this question, we have considered the generation of magnetic field by the flow of a thermally convecting Boussinesq fluid contained in a  rotating spherical shell, thus modeling stellar or planetary core configurations. Fixed temperatures are imposed at both inner and outer boundaries, no-slip boundary conditions are used for the velocity field, and both boundaries are electrically insulating. The dimensionless parameters are the shell aspect ratio $\gamma=r_i/r_o$, the  magnetic Prandtl number $Pm=\nu/\eta_0$, the Ekman number $E=\nu/\Omega D^2$ and the Rayleigh number $Ra=\alpha g_0 \Delta T D/(\nu\Omega)$, where $D=r_o-r_i$ is the gap and $\Omega$, $\nu$, $\eta_0$, $\kappa$, $\alpha$ and $g_0$ are, respectively, the rotation rate, the kinematic viscosity, the spatially averaged magnetic
diffusivity, the thermal diffusivity, the thermal expansion coefficient, and the gravity at the outer sphere. Equations of magnetohydrodynamics for the velocity $v$, magnetic field $B$ and temperature $T$ are solved with the help of the code PaRoDy \cite{Dormy98}, which has been modified to take into account the spatial variation of the  electrical conductivity.
 As a simple example, we assume here that the magnetic diffusivity $\eta$ depends on the temperature as $\eta=\eta_0+k(T-T_0)$, where the proportionality coefficient $k$ is kept as a control parameter.  Several configurations have been considered: conductivity depending only on the temperature fluctuations or on both the temperature fluctuations and  the background temperature profile. In addition, different widths of the spherical shell have been tested.  Note that effects of radially varying conductivity were studied in \cite{jones}, in which it was shown that a low-conductivity layer close to the core-mantle boundary may explain Mercury's weak observable magnetic field. Here, we rather study the case of conductivity depending on the temperature field that can fluctuate in all directions.

\begin{figure}[htb!]
\begin{center}
\includegraphics[width=90mm]{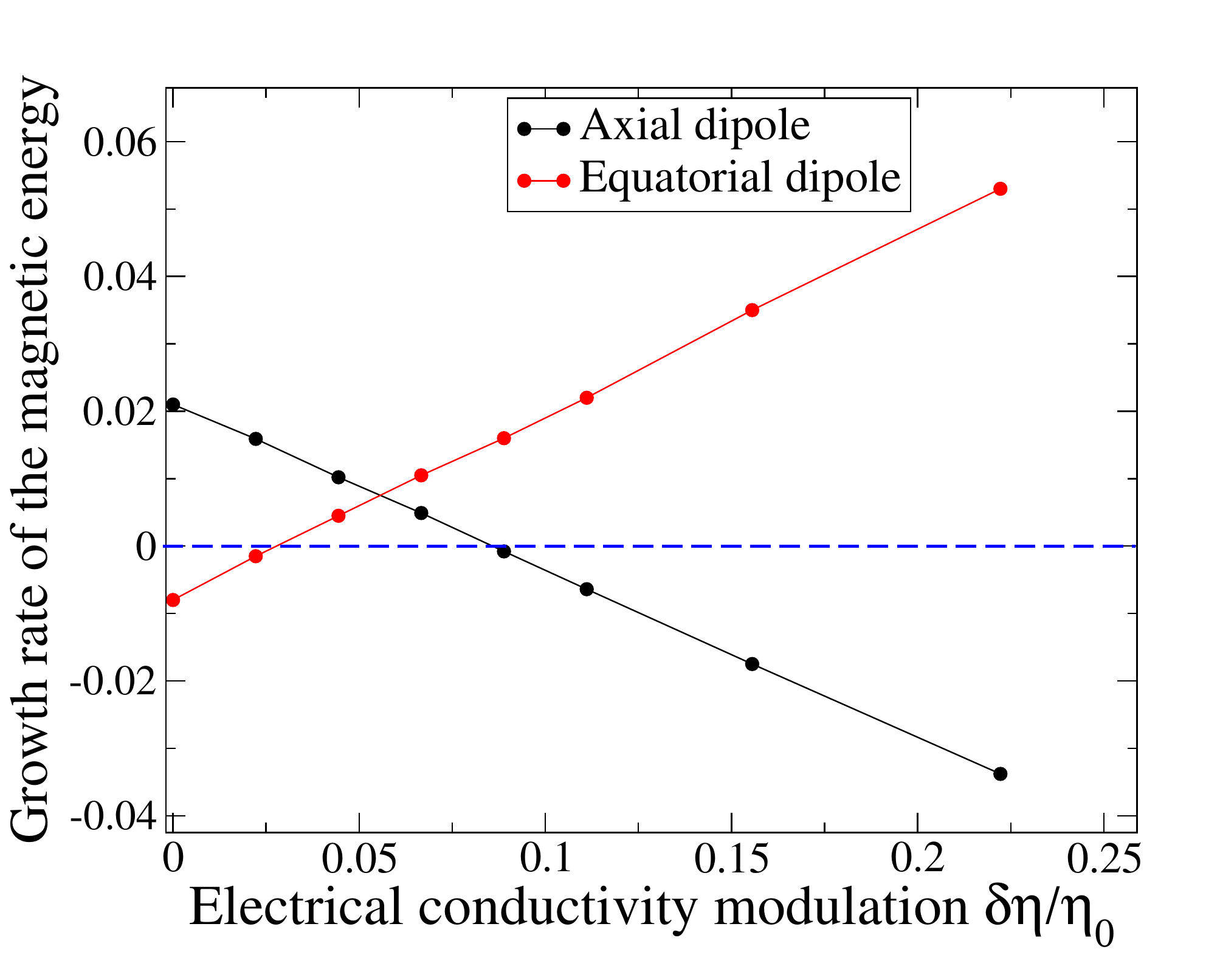}
\caption{Growth rate of the magnetic energy in the kinematic phase as a function of the (temperature-driven) electrical conductivity modulation in a dynamo simulation, for $E=6.10^{-4}$, $Pm=7.9$ and $Ra/Ra_c=2.2$, for axial (black) and equatorial (red) dipole modes. Note that the axial dipole obtained at $\delta\eta/\eta_0=0$ is replaced by a transverse dipole in presence of conductivity modulation.}
\label{fichristophe1}
\end{center}
\end{figure}

Although the  parameter space is huge and further work is required to fully characterize the effect of a varying conductivity, it can be  identified that a transverse dipolar field  benefits from a modulation of electrical conductivity in typical geodynamo simulations. Fig. \ref{fichristophe1} shows the growth rate of the magnetic field as a function of the amplitude of the conductivity modulation for $\gamma=0.35$, $E=6.10^{-4}$, $Ra/Ra_c=2.2$ ($Ra=123$) and $Pm=7.9$. In the case of an homogeneous conductivity ($\delta\eta/\eta_0=0$) , an axial dipole is observed (black curve), as usual for these parameters. As the coupling coefficient $k$ between the temperature and the conductivity is increased, the growth rate of this dipole decreases until it becomes kinematically stable. In contrast, the growth rate of the equatorial dipole mode (red curve) increases from negative to positive values. As soon as $\delta \eta/\eta_0$ reaches $5 \%$, the modulation of the electrical conductivity   changes the structure of the dynamo field, replacing the axial dipole by a transverse one. For both modes, we observe   a linear relation between the growth rate and the conductivity modulation,  as predicted by our theory. 

\begin{figure}[htb!]
\begin{center}
\includegraphics[width=90mm]{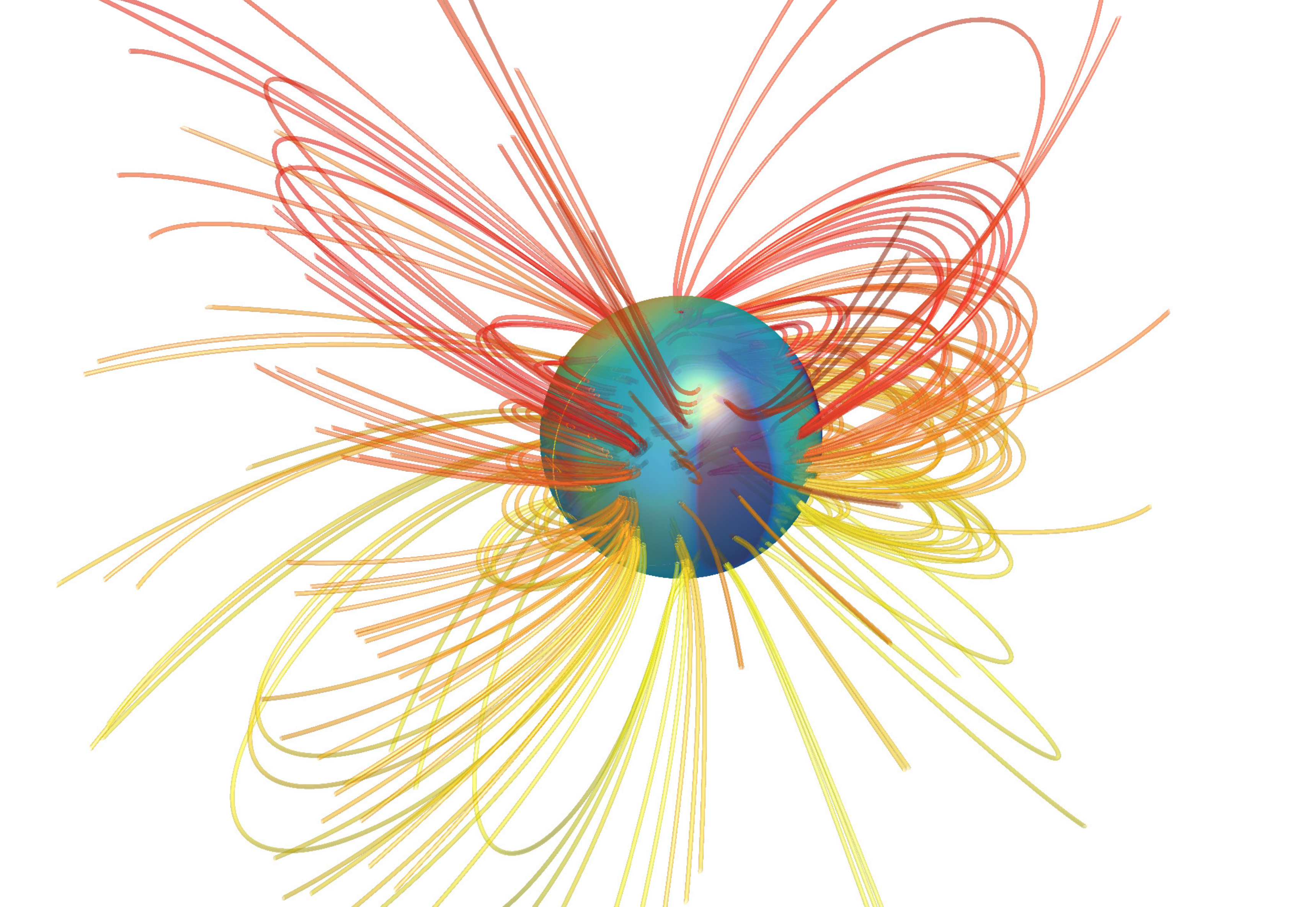}
\caption{Structure of the saturated equatorial dipole generated for $E=10^{-3}$, $Pm=7$ , $Ra/Ra_c=2$ and $\delta\eta/\eta_0=0.4$. The colored sphere indicates amplitude of the radial magnetic field at the surface of the core-mantle boundary and magnetic field lines in the insulating mantle are shown. } 
\label{fichristophe2}
\end{center}
\end{figure}

This effect of the conductivity modulation is observed in a wide region of the parameter space. For instance, Fig. \ref{fichristophe2} displays the spatial structure of the dynamo magnetic field obtained for $E=10^{-3}$, $Ra/Ra_c=2$, $Pm=7$ and $\delta\eta/\eta_0=0.4$, corresponding to an equatorial dipolar field. Note that for these values of $E$ and $Ra$, the conductivity fluctuations decrease the dynamo onset by roughly $20\%$ compared to the homogeneous case.

\begin{figure}[htb!]
\begin{center}
\includegraphics[width=85mm]{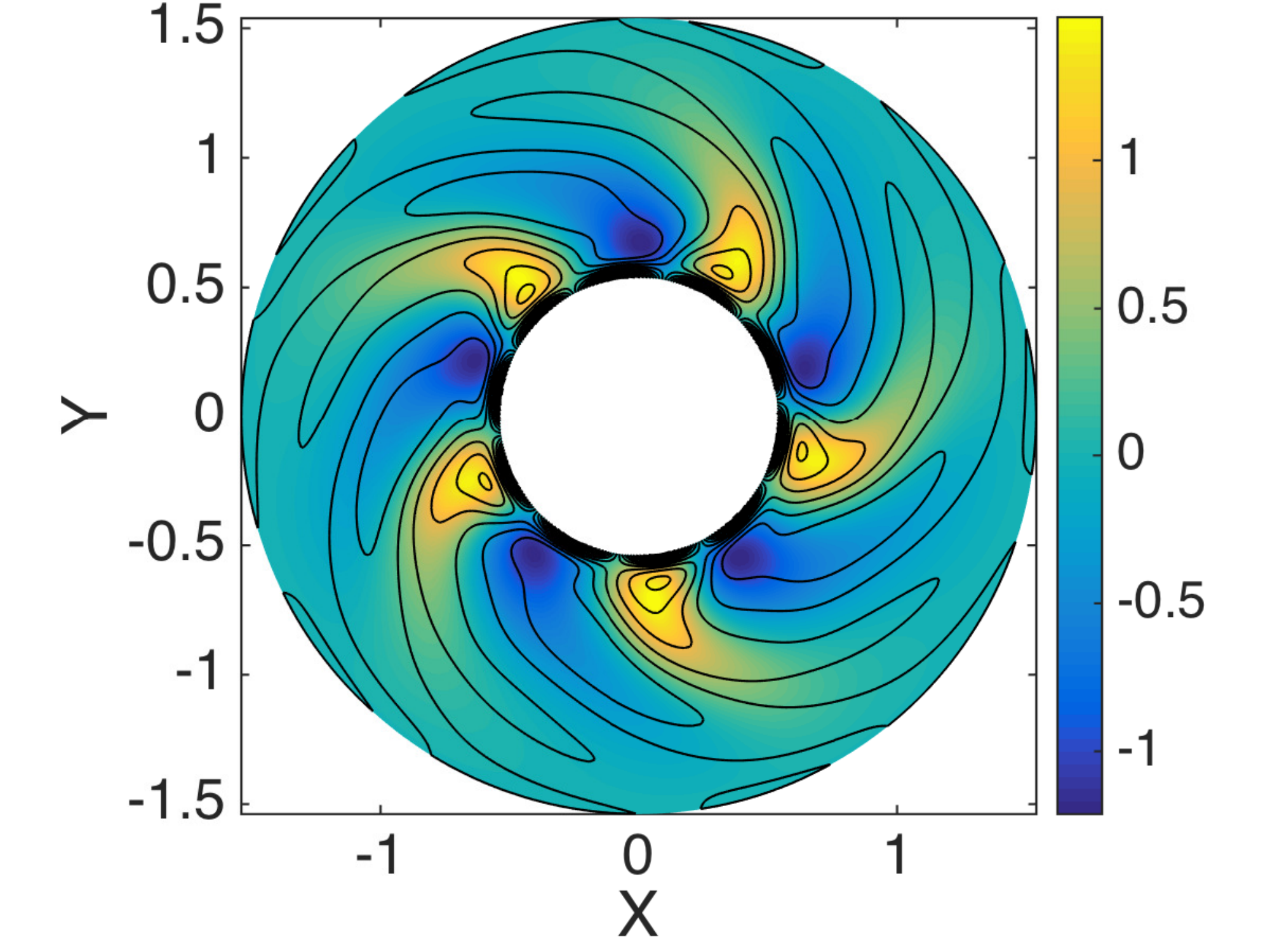}
\caption{Equatorial cut of the purely hydrodynamical state obtained for $E=6.10^{-4}$ and $Ra/Ra_c=2.2$. Colorplot displays the amplitude of the azimuthal temperature gradient $\partial_\phi T$ whereas black lines are isocontours of the axial component of the vorticity ${(\bf \nabla \times u}).{\bf e_z}$. Note the strong correlation between the two quantities.}
\label{fichristophe3}
\end{center}
\end{figure}

To understand how such a temperature-dependent conductivity decreases the dynamo onset, it is important to note that geophysical flows, strongly affected by Taylor-Proudman theorem, mainly consist of several columnar vortices arranged along the azimuthal direction (the so-called Busse columns \cite{bussecol}) with temperature gradient maximum at the center of the vortices.  This convective pattern is therefore characterized by a strong correlation between the axial vorticity and the azimuthal gradient of temperature, as illustrated in  Fig.\ref{fichristophe3}. The component $({\bf \nabla\times u})|_z  . {\bf \nabla}_\phi (\delta\eta)$ is mainly localized in the equatorial plane, thus suggesting that this non-diagonal term of the $\alpha^{\sigma}$-tensor is responsible for the generation of the field. Note that  this differs from the diagonal part of the usual $\alpha$-effect, which vanishes in the equatorial plane. The $\alpha^{\sigma}$-effect, being  strong in the equatorial plane,  thus provides a possible explanation for the equatorial dipolar component of the  magnetic field observed in Neptune and Uranus \cite{uranusobs,uranus}.

To discuss further the relevance of this  effect, it is interesting to compare its efficiency with the one of an $\alpha^2$-dynamo. In scale separation, the onset for an $\alpha^2$ dynamo is given by $V \sqrt{l L}/\eta=C_1$ where $C_1$ is a constant, $V$ the amplitude of the velocity, $l$ the wavelength of the flow and $L$ the size over which the large scale field varies. For an  $\alpha^{\sigma}$-dynamo, the onset is $\delta V L/\eta=C_2$ where $C_2$ is a constant and $\delta$ the amplitude of the relative variations of conductivity. Thus, for a flow that is prone to both effects, the $\alpha^{\sigma}$-dynamo leads to a smaller onset provided $\delta \sqrt{\frac{L}{l}} \gg 1$,  meaning that this new kind of dynamo is expected when scale separation is large enough. 

As the efficiency of the $\alpha^{\sigma}$-effect depends crucially on the variations of the electrical conductivity, it is  worth discussing   possible sources for these variations that are met in nature. 
%In a telluric planet such as the Earth, the electrical conductivity greatly varies with the depth in the liquid core due to the increase of temperature and pressure. Variations of the order of $20 \%$ are estimated \cite{gomi}. 
%In addition to these static radial variations, one has to consider the effect of the convective temperature fluctuations that are the sources %of both the conductivity variations and the velocity fluctuations.   
 In a telluric planet such as the Earth, the time-averaged electrical conductivity varies with the depth in the liquid core due to the increase of temperature and pressure \cite{gomi}. However, one has to consider the effect of the convective temperature fluctuations which are quite smaller than the static radial variations. These fluctuations are the sources of both the conductivity variations and the velocity fluctuations, and simple estimates of their intensities
 show that the efficiency of the  $\alpha^{\sigma}$-effect is larger than the one of the  usual $\alpha$-effect when scale-separation is large enough.  It is then worth noting that rapid rotation results in a drastic shortening of the characteristic length-scale of convective pattern \cite{RB}, so that this new kind of dynamo should be relevant for rapidly rotating astrophysical objects.

In the case of the Sun, temperature differences of $200-400 K$ are measured at the surface between ascending and descending plumes. For  linear dependance of $\sigma$ on $T$, this would correspond to relative variations of $\sigma$ of $3$ to $7 \%$, making the dimensionless parameter $\delta V L/ \eta$  large enough for the  $\alpha^{\sigma}$-effect to play a role.

The magnetic field is known to play  a role in the dynamics of the Sun convective zone. This sheds light on  another possible source for conductivity variations:  Ohmic dissipation itself.
On can imagine that the electric currents heat up locally the fluid so that it modifies the conductivity and affects the efficiency of the $\alpha^{\sigma}$-effect. This would result in a non-linear mechanism that could act as a saturation mechanism if the efficiency of the effect is decreased by Joule heating or, if the efficiency is increased,  could be responsible for a non-linear amplification.  This effect thus provides a new scenario for  a subcritical dynamo instability.

In the laboratory, the $\alpha^{\sigma}$-effect can be used to build dynamo flows simpler than those considered so far. Indeed, the possibility to use  planar flow greatly simplifies the geometrical constraints. Using  liquid sodium which displays a decrease of conductivity of more than $25 \%$ between $100$ and $200$ degrees, a periodic array of counter-rotating vortices with proper control of temperature variations would generate a dynamo at a magnetic Reynolds number achievable at the laboratory scale.

Finally, one may use the $\alpha^{\sigma}$-effect to modify the onset of an existing laboratory dynamo set up.  The Karlsruhe dynamo \cite{karls} is the simplest configuration to analyze as it is made of a periodic array of helical flows. By imposing conductivity variations between the different vortices, an $\alpha^{\sigma}$-effect is added to the $\alpha$-effect. A corresponding decrease of the critical magnetic Reynolds number proportional to $\delta \eta/(Vl)$ is expected, leading to a possible threshold reduction of roughly  $10 \%$.

{\bf Acknowledgements:}

We thank S. Fauve for suggesting to perform the scale-separation calculation and for several discussions. F. P. thanks T. Alboussi\`ere, K. Ferri\`ere and R. Raynaud for fruitfull discussions.  This work was granted access to the HPC resources of MesoPSL financed   by the Region Ile de France and the project Equip@Meso (reference
ANR-10-EQPX-29-01) of the programme Investissements d'Avenir supervised by the Agence Nationale pour la Recherche.


\begin{thebibliography}{0}
\bibitem{larmor} Larmor, J., Reports of the British Association, {\bf 87}, 159-160, 1919.
\bibitem{2D} Zeldovich Y., Ruzmaikin A., Sov. Phys. JETP 51 (3) 1980.
\bibitem{moff}  Moffatt, H. K., {\it Magnetic Field Generation in Electrically Conducting Fluids}, 1978, (Cambridge University Press).
\bibitem{busseandwicht} Busse F. and Wicht J., Geophys.
Astrophys. Fluid Dyn. 64, 135-144, 1992.
\bibitem{gallet} This is discussed in detail in the similar situation of a varying magnetic permeability by B. Gallet et al., Europhysics Letters, 
97, 69001, 2012  and B. Gallet et al., J. Fluid Mech.,  727, 161-190, 2013.
\bibitem{Dormy98} Dormy E., Cardin P. and Jault D.,
Earth Planet. Sci. Lett. 160, 15, 1998.
\bibitem{jones} Gomez-Perez, N. et al., PEPI 181, 42-43, 2010.
\bibitem{bussecol} Busse, F.H., J. Fluid Mech. 44, 441-460, 1970.
\bibitem{uranusobs} Ness, N. F. et al., Science 233, 85–89, 1986. Ness, N. F. et al.,  Science 246, 1473-1478, 1989.
\bibitem{uranus} For other possible explanations of the equatorial component of the   iced giant  magnetic field,  see  Stanley, S. and Bloxham, J., Nature, 428, 6979, 2004. J. Aubert, J. Wicht
EPSL 221, 409-419, 2004. Gissinger C. et al.,  Phys. Rev. Lett., 108, 234501, 2012.
\bibitem{gomi} Gomi H. et al., PEPI (224), 88-103, 2013.
\bibitem{RB} Chandrasekhar, S.,  {\it Hydrodynamic and hydromagnetic stability}, 1961, (Clarendon Press, Oxford).
\bibitem{karls} Stieglitz R. and M\"uller U., Phys. Fluids, {\bf 13}, 561 (2001). 
\end{thebibliography}
\end{document}